\documentclass[10pt,conference,a4paper]{IEEEtran}

\pdfpageattr{/Group << /S /Transparency /I true /CS /DeviceRGB >>}

\usepackage[utf8]{inputenc}
\usepackage[T1]{fontenc}
\usepackage{textcomp}
\usepackage{microtype}
\usepackage{calc}
\usepackage[normalem]{ulem}

\usepackage{lipsum}

\usepackage[range-phrase=--,per-mode=symbol-or-fraction,binary-units=true,range-units=single,list-units=single,detect-all,forbid-literal-units=true]{siunitx}
\AtBeginDocument{
\DeclareSIUnit\bit{bit}
\DeclareSIUnit\byte{Byte}
\DeclareSIUnit\decibelm{dBm}
\DeclareSIUnit\vehicle{veh}
}

\usepackage{silence}
\usepackage{csquotes}
\usepackage{amsmath}
\usepackage{amssymb}
\usepackage{amsfonts}
\usepackage{amsthm}

\usepackage{mathtools}

\usepackage{booktabs}
\usepackage{url}

\usepackage[inline]{enumitem}

\usepackage[caption=false,font=footnotesize]{subfig}
\usepackage[pdftex]{graphicx}
\DeclareGraphicsExtensions{.pdf,.png,.jpg}
\usepackage{tikz}
\usetikzlibrary{arrows}
\usetikzlibrary{calc}
\usetikzlibrary{chains}
\usetikzlibrary{scopes}

\usepackage[american]{babel}
\usepackage{hyphenat}

\usepackage{algorithmic}
\usepackage{algorithm}

\usepackage[capitalize,noabbrev,nameinlink]{cleveref}

\RequirePackage{xstring}
\RequirePackage{xparse}
\RequirePackage[]{acro}
\makeatletter
\@ifpackagelater{acro}{2019/02/17}{
	\NewDocumentCommand\acrodef{mO{#1}mG{}}{\DeclareAcronym{#1}{short={#2}, long={#3}, foreign-plural={}, #4}}
}{
	\NewDocumentCommand\acrodef{mO{#1}mG{}}{\DeclareAcronym{#1}{short={#2}, long={#3}, #4}}
}
\makeatother
\acrodef{AIFS}{Arbitration Inter-Frame Spacing}
\acrodef{AoI}{Age of Information}
\acrodef{AWGN}{Additive White Gaussian Noise}
\acrodef{BS}{Base Station}
\acrodef{BSS}{Basic Service Set}
\acrodef{CAM}{Cooperative Awareness Message}
\acrodef{CPM}{Collective Perception Message}
\acrodef{CBF}{Contention-Based Forwarding}
\acrodef{CBR}{Channel Busy Ratio}
\acrodef{CDF}{Cumulative Distribution Function}
\acrodef{CCDF}{Complementary Cumulative Distribution Function}
\acrodef{CDMA}{Code Division Multiple Access}
\acrodef{CCH}{Control Channel}
\acrodef{CSMA}{Carrier-Sense Multiple Access}
\acrodef{C-ITS}{Cooperative Intelligent Transportation System}{short-plural-form={C-ITS}}
\acrodef{DENM}{Decentralized Environmental Notification Message}
\acrodef{DIFS}{DCF Inter-Frame Space}
\acrodef{DSRC}{Dedicated Short-Range Communication}
\acrodef{DTMC}{Discrete Time Markov Chain}
\acrodef{FCFS}{First Come First Served}
\acrodef{FDMA}{Frequency Division Multiple Access}
\acrodef{GFMA}{Grant-Free Multiple Access}
\acrodef{IoT}{Internet of Things}
\acrodef{ITS}{Intelligent Transportation System}{long-plural-form={}}
\acrodef{IVC}{Inter Vehicle Communication}
\acrodef{LCFS}{Last Come First Served}
\acrodef{LCFSwO}{Last Come First Served with Overwrite}
\acrodef{LDM}{Local Dynamic Map}
\acrodef{M2M}{Machine-to-Machine}
\acrodef{MAC}{Medium Access Control}
\acrodef{MCM}{Maneuver Coordination Message}
\acrodef{MIMO}{Multiple Input Multiple Output}
\acrodef{MPR}{Multi-Packet Reception}
\acrodef{MTC}{Machine Type Communications}
\acrodef{NOMA}{Non-Orthogonal Multiple Access}
\acrodef{OBU}{On-Board Unit}
\acrodef{OFDMA}{Orthogonal Frequency Division Multiple Access}
\acrodef{OMA}{Orthogonal Multiple Access}
\acrodef{PDF}{Probability Density Function}
\acrodef{PDR}{Packet Delivery Ratio}
\acrodef{RSU}{Road Side Unit}
\acrodef{SA}{Slotted ALOHA}
\acrodef{SCH}{Service Channel}
\acrodef{SIC}{Successive Interference Cancellation}
\acrodef{SNR}{Signal to Noise Ratio}
\acrodef{SNIR}{Signal to Noise plus Interference Ratio}
\acrodef{V2I}{Vehicle-to-Infrastructure}
\acrodef{V2V}{Vehicle-to-Vehicle}
\acrodef{V2X}{Vehicle-to-Everything}
\acrodef{VANET}{Vehicular Ad Hoc Network}
\acrodef{VLC}{Visible Light Communication}
\acrodef{TDMA}{Time Division Multiple Access}
\acrodef{UAV}{Unmanned Aerial Vehicle}
\acrodef{URLLC}{Ultra-Reliable Low-Latency Communications}

\usepackage[color]{changebar}
\def\todoCtd#1{%
	TODO: #1%
	\ifx&#1&...\fi%
	\endgroup
	\cbend
	\relax
}

\NewDocumentCommand\IEEE{ s m >{\SplitArgument{4}{/}}d[] }{%
    \IfBooleanTF{#1}{}{IEEE\,}% suppress IEEE when using starred form
    \nolinebreak[2]% this is a somewhat bad place for a line break
    #2%
    \IfNoValueTF{#3}{%
    }{%
        \sommerIEEELettersSlashed#3%
    }%
}
\newcommand{\sommerIEEELettersSlashed}[5]{%
    \IfNoValueTF{#2}{%
    }{%
        \nolinebreak[3]% multiple letters, this is just a very bad place for a line break
    }%
	#1%
	\IfNoValueTF{#2}{}{/#2}%
	\IfNoValueTF{#3}{}{/#3}%
	\IfNoValueTF{#4}{}{/#4}%
	\IfNoValueTF{#5}{}{/#5}%
}

\begin{document}
\title{SIC-based Random Multiple Access Protocol: Fixed or Adaptive Approach\\
\thanks{This work was partially supported by the European Union under the Italian National Recovery and Resilience Plan (NRRP) of NextGenerationEU, partnership on “Telecommunications of the Future” (PE0000001 - program “RESTART”).}
}
% Optimizing the AoI-energy trade-off in massive IoT with SIC multiple access

\author{%
\IEEEauthorblockN{%
	Asmad Razzaque\IEEEauthorrefmark{1}%
	~and
	Andrea Baiocchi \IEEEauthorrefmark{1}%
}%

\IEEEauthorblockA{
	\IEEEauthorrefmark{1}Dept.\ of Information Engineering, Electronics and Telecommunications (DIET), University of Rome Sapienza, Italy
}%

%\IEEEauthorblockA{
%	\IEEEauthorrefmark{2}Interdisciplinary Centre for Security, Reliability and Trust (SnT), University of Luxembourg
%}%

%
\texttt{%
	%\{%
	\{asmadbin.razzaque,andrea.baiocchi\}%
	%\}
	@uniroma1.it%
}%
%,
%\texttt{%
%	%\{%
%	ion.turcanu%
%	%\}
%	@uni.lu%
%}%
}

\maketitle

\begin{abstract}\nohyphens{%
Efficient data collection from a multitude of \ac{IoT} devices is crucial for various applications, yet existing solutions often struggle with minimizing access delay and \ac{AoI}, especially when managing multiple simultaneous transmissions and access strategies.
This challenge becomes increasingly critical as \ac{IoT} deployments continue to expand, demanding robust mechanisms for handling diverse traffic scenarios.
In this study, we propose a novel approach leveraging \ac{SIC} based on adaptive and fixed parameter schemes to address these limitations.
By analyzing both throughput and \ac{AoI} along with access delay, we demonstrate the effectiveness of our adaptive approach compared to the fixed approach, particularly in scenarios featuring heavy and light traffic.
Our findings highlight the pivotal role of adaptive approaches in optimizing data collection processes in \ac{IoT} ecosystems, with a particular focus on minimizing access delay, \ac{AoI}, and spectral efficiency.
}\end{abstract}


\acresetall
\IEEEpeerreviewmaketitle

% -------------- Section end marker --------------
%                _       _
%               ( )_    ( )
%    ___  _   _ | ,_)   | |__     __   _ __   __
%  /'___)( ) ( )| |     |  _ `\ /'__`\( '__)/'__`\
% ( (___ | (_) || |_    | | | |(  ___/| |  (  ___/
% `\____)`\___/'`\__)   (_) (_)`\____)(_)  `\____)
%
% -------------- Section end marker --------------

\section{Introduction}
\label{sec:introduction}

The rise of the \ac{IoT} has sparked a big change in how we use technology. 
It is affecting everything from healthcare and education to industry and transportation \cite{Guo2021}.
As IoT applications evolve towards unprecedented realms, the concept of massive access emerges as a crucial paradigm aimed at facilitating efficient and dependable communication for a wide range of IoT devices.
Characterized by low power consumption, massive connectivity, short packet transmission, and minimal signaling overhead, the requirements of massive access herald the vision of a data-driven society envisaged by 6G, wherein instantaneous and boundless connectivity is extended to a multitude of entities, ranging from static sensors to autonomous devices, irrespective of time and location.

Within the context of massive multiple access in \ac{IoT}, the challenge lies in ensuring accurate and efficient data transmission from a vast number of devices, especially considering their sporadic activity across the network.
Addressing these challenges necessitates innovative multiple-access techniques that offer massive connectivity and low latency to future IoT systems.
One promising solution is grant-free random access, wherein each active device directly transmits data to the base station without contention using orthogonal resources, which reduces the signaling overhead significantly \cite{LuiPetar2018}. 

However, relying exclusively on orthogonal resources has its limitations, especially in large-scale \ac{IoT} setups.
\ac{NOMA} emerges as a potential solution, offering two primary categories: code-domain \ac{NOMA} and power-domain \ac{NOMA}. 
Power-domain \ac{NOMA}, exemplified by its implementation in 3GPP LTE, involves superposing power from multiple users at the transmitter, with successive interference cancellation (SIC) employed at the receiver for decoding \cite{Saito2013,Dai2015,DingChoi2017}.
Unlike conventional techniques such as \ac{CDMA} and orthogonal multiple access methods, \ac{NOMA} leverages the power domain for user multiplexing, thereby enhancing spectral efficiency and accommodating diverse connectivity requirements prevalent in massive IoT deployments.

Recent progress in grant-free multiple access has seen remarkable advancements in adopting \ac{NOMA}.
These advancements utilize techniques spanning from deep learning to reinforcement learning to enhance user detection accuracy, spectral efficiency, and interference mitigation \cite{Emir2021,Fayaz2021}.
However, the realization of scalable multi-packet reception systems, capable of accommodating thousands of transmitting nodes, remains a challenge.
While \ac{SIC} offers performance gains, the benefits diminish with each iteration, highlighting the need for further research and optimization in this domain \cite{Zanella2012}.
It has been seen that \ac{SIC} enhances the connectivity in short packet data transmission as the number of users increases \cite{Razzaque2022,LinDai2018}.
This calls for an efficient adaptive protocol that can ensure necessary connectivity with minimal signaling overhead and reduce the age of information which is a parameter to measure the freshness of information.
Such a protocol should dynamically adjust transmission parameters as the number of users fluctuates in the highly sporadic IoT environment.
An adaptive parameter-based random access protocols have been proposed recently \cite{SHASHIN2023,Gizik2021,Jeon2023,ITC35} to improve the system capacity and low latency.

However, there exists a significant gap in the literature regarding the understanding of the role of decoding capabilities given by SIC toward different transmission schemes.
This scenario becomes critical in the vast world of IoT, where every device is constantly sending and receiving data, and ensuring that information flows smoothly.
The conventional methods, as explained struggle a lot to maintain the flow of traffic in a reliable way.
To tackle this challenge, we are exploring the \ac{SIC}, which has the potential to revolutionize how we handle data in \ac{IoT} networks by allowing us to decode multiple data streams even when they overlap.
%This means we can potentially handle more data simultaneously and with greater efficiency.

%In this paper, we aim to understand the impact of \ac{SIC} based optimal parameters \cite{ITC35} toward the non-saturated model.
%We have investigated how much gain the adaptive scheme brings while using the optimal parameters, in massive \ac{IoT} scenarios.
%How much do we lose in terms of gain if the system doesn't adapt the parameters to a heavy traffic scenario especially.
%Is this loss acceptable in terms of system complexity that adaptivity brings in

%We define a general model of multiple access, consisting of $N$ non-saturated nodes that generate update messages sent to a central collecting node, equipped with a \ac{SIC} receiver.
%\ac{SIC} is modeled in a general way, not tied to specific implementations, we focus our attention toward adaptive and fixed parameters which are driven based on SIC decoding capabilities \cite{ITC35}

%\textcolor{blue}{Here I have to add a few more strong point to end this paragraph well and explain a bit after adding numerical section that how we are comparing adaptive and fixed parameters strategy}

The real twist comes while considering a non-saturated network to investigate how we can tweak and optimize the network with the help of \ac{SIC} based parameters to work even better in different traffic scenarios with lower complexity, as \ac{SIC} requires an adaptive system to enhance the performance gains based on the transmitting nodes.
Hence, we investigate two different strategies: adaptive and fixed parameter schemes.
The adaptive approach is like having a smart system that can adjust its settings based on what is happening in the network.
So, if there is a sudden surge in data traffic, the system can automatically adapt to handle it more effectively.
On the other hand, the fixed parameter scheme sticks to a set of predetermined settings regardless of the network conditions.
While this approach may be simpler, it might not always be the most efficient, especially when the network experiences fluctuations in traffic.
We have set the optimal parameters toward the fixed and adaptive parameter schemes with \ac{SIC} provided gains in terms of throughput, and age of information as given in \cite{ITC35,ANTE2023}.

We seek to gain insights into the trade-offs inherent in the implementation by exploring the two strategies. 
How much does the adaptive approach outperform the fixed one, and is the increased complexity of adaptivity justified in terms of overall system performance compared to fixed parameters?
We delve into the systems success rate, access delay, throughput, and \ac{AoI} to address these questions, offering insights that will ease this revolution of IoT toward future multiple access schemes.
This study highlights the balance between simplicity and efficiency, ensuring smooth data flow even in the busiest IoT environments.

%A fixed grant-free scheme is defined, where system parameters are selected to optimize the achieved sum rate as the
%We believe that this work opens up several paths for further research. 
%These include implementing practical algorithms to realize the promises assessed by the model.
%Defining distributed algorithms to adapt multiple access parameters, inspired by the insights gained from the presented mode is another research line we are pursuing.

The rest of the paper is organized as follows.
\cref{sec:model_and_analysis} introduces the analytical model along with system analysis, including assumptions, definitions, and notation.
Numerical results are reported in \cref{sec:num_eval}.
Finally, conclusions are drawn in \cref{sec:conclusions}.

% -------------- Section end marker --------------
%                _       _
%               ( )_    ( )
%    ___  _   _ | ,_)   | |__     __   _ __   __
%  /'___)( ) ( )| |     |  _ `\ /'__`\( '__)/'__`\
% ( (___ | (_) || |_    | | | |(  ___/| |  (  ___/
% `\____)`\___/'`\__)   (_) (_)`\____)(_)  `\____)
%
% -------------- Section end marker --------------

\section{System Model and Analysis}
\label{sec:model_and_analysis}

Model assumptions and notation are introduced in \cref{subsec:modeldef}.
Analysis of the model is carried out in \cref{subsec:modelanalysis}.

\subsection{Model definition}
\label{subsec:modeldef}
We consider a network of $n$ nodes, sending update messages to a sink, referred to as \ac{BS}.
The time axis is slotted.
A transmission attempt is made by backlogged nodes in each slot with a given probability.
% More in-depth, let $Q(t)$ denote the number of backlogged nodes at the beginning of the slot $t$.
More in-depth, a backlogged node attempts transmission with probability $p$ and picks its modulation and coding scheme according to a target \ac{SNIR} $\gamma$.
Assuming an \ac{AWGN} communication channel, the target \ac{SNIR} $\gamma$ is tied to the achievable spectral efficiency $\eta$ (bit per symbol) of the adopted modulation and coding scheme according to $\eta = \log_2(1+\gamma)$.
A packet is correctly decoded if its average \ac{SNIR} at the \ac{BS} is no less than $\gamma$.
We assume also that feasible values of $\gamma$ are upper limited to some value $\gamma_{\text{max}}$, related to maximum available transmission power and target coverage distance of the \ac{BS}.

Let $L$ be the length of the transmitted packets and $W$ be the channel bandwidth.
It is assumed that the slot size just fits a packet transmission time.
The time required to transmit a packet for a given target \ac{SNIR} $\gamma$ is
\begin{equation}
\label{eq:slottimeduration}
T(\gamma) = \frac{ L }{ W \log_2(1+\gamma) }
\end{equation}

We consider two approaches to set transmission parameter $p$ and $\gamma$.
First, the fixed transmission parameter setting, in which $p = p^*$ and $\gamma = \gamma^*$ irrespective of the actual number of backlogged nodes in each time slot.
In this case, the slot size is fixed and equal to $T^* = T(\gamma^*)$ from \cref{eq:slottimeduration}, even if no nodes transmits.
Second, the Adaptive transmission parameter setting, where $p = p(t)$, and $\gamma = \gamma(t)$ are set as a function of the number $Q(t)$ of backlogged nodes at the beginning of slot $t$.
In this case, the slot size varies with the number of backlogged nodes.
More in depth, we set $p = p^*_k$ and $\gamma = \gamma^*_k$, if $Q(t) = k$, for $k \ge 1$.
In case $Q(t) = 0$, the slot size is assigned a fixed value $T_0$, that does not depend on values of $p$ and $\gamma$.

New messages are generated at each node according to a Poisson process with mean rate $\lambda$.
Messages are generated in upper layers and passed down to the MAC layer entity.
Once a node MAC entity is engaged with contention/transmission of a message, it cannot be interrupted.
If the MAC entity of the node is engaged in contention/transmission of a message, a new arriving message is dropped. 
% This is consistent with the update message collection scenario, where only the latest update is relevant to applications.
% It is shown in \cite{Eenennaam2011} that dropping older messages is beneficial to delay. 
It is shown in \cite{Baiocchi2021ITC33} that having no buffer at the MAC level is beneficial to \ac{AoI}, which is the relevant metric in the considered use case of update messages.
% The reason is that keeping new messages extends the time that a node is backlogged and contends with others for channel resources. 
% The resulting higher level of contention affects adversely \ac{AoI}, setting back the potential advantage of maintaining the latest message in a buffer.
% The performance benefit applies not only to mean values, but also to tails of probability distributions of the \ac{AoI}. 

The transmission power level is adjusted to compensate for path loss and slow fading.
Then, the received power level is modeled as $P_{\text{rx}} = G_f P_0$, where $G_f$ is the fast fading gain, characterized as a negative exponential random variable with mean 1 (Rayleigh fading), and $P_0$ is the average received power level at \ac{BS}. 
$P_0$ is set to a target value, so that the probability of failing to decode a packet sent by a \emph{single} transmitting node is no more than $\epsilon$.
% The path gain of the communication channel is modeled as $G = G_d G_s G_f$, where $G_d$ is the deterministic gain accounting for distance between the transmitting node and the \ac{BS}, $G_s$ is the shadowing gain accounting for obstacles, and $G_f$ is the fading gain accounting for multi-path.
% We assume that the first two components do not vary over time, while the third one is sampled independently slot by slot from a negative exponential \ac{PDF} with mean 1 (Rayleigh fading).
% The transmission power level $P_{\text{tx}}$ is adjusted to compensate for the first two components.
%Therefore, the average received power level $P_0$ at \ac{BS} is set to a target value, so that the probability of failing to decode a packet sent by a single transmitting node is no more than $\epsilon$.
Decoding of a packet sent by a single transmitting node is successful, if the \ac{SNR} exceeds $\gamma$, i.e., $P_{\text{rx}} / P_N \ge \gamma$, where $P_N$ is the background noise power level. 
Since the average received power level is $P_0$, the requirements translates to
\begin{equation}
\label{ }
\frac{ G_f P_0 }{ P_N } \ge \gamma \qquad \text{w.p. } 1-\epsilon
\end{equation}
Given that $G_f$ has a negative exponential \ac{PDF}, $P_0$ is set so that $\mathcal{P}( G_f P_0 / P_N \ge \gamma ) = e^{ - \gamma P_N/P_0} = 1-\epsilon$.
Hence, the target \ac{SNR} level $S_0$ at the receiving \ac{BS} is set as follows:
\begin{equation}
\label{eq:targetaveragerxSNR}
S_0 = \frac{ P_0 }{ P_N } = \frac{ \gamma }{ -\log(1-\epsilon) } = \frac{ \gamma }{ c }
\end{equation}
where we have introduced the constant $c = -\log(1-\epsilon)$.

We assume an ideal \ac{SIC} receiver.
Let $h$ packets be received simultaneously in the same slot and let $S_j, \, j = 1,\dots,h$ be their respective received power levels, normalized with respect to the background noise power level\footnote{Note that $S_j = G_{f,j} S_0$, where $G_{f,j}$ is the fading path gain of the $j$-th user and $S_0$ is given in \cref{eq:targetaveragerxSNR}.}.
Let the $S_j$'s be ordered in descending order, i.e., $S_1 \ge S_2 \dots \ge S_h$ (ties are broken at random).
The \ac{SIC} receiver works as follows.
Provided decoding of packets $1,\dots,\ell-1$ be successful, packet $\ell$ is decoded successfully if and only if the following inequality holds:
\begin{equation}
    \frac{ S_\ell }{ 1 + \sum_{ r = \ell+1 }^{ h }{ S_r } } \ge \gamma
\end{equation}
Note that we assume perfect interference cancellation.
Hence, the residual interference is due only to signals weaker than the $\ell$-th one.

%For comparison purposes, we will also consider a receiver not endowed with \ac{SIC} capability.
%In that case, multi-packet reception is still possible, thanks to capture effect.
%Specifically, packet $\ell$ is successfully received if and only if its \ac{SNIR} exceeds $\gamma$, which in case of capture only turns into the following inequality:
%\begin{equation}
%    \frac{ S_\ell }{ 1 + \sum_{ r = 1, r \ne \ell }^{ h }{ S_r } } \ge \gamma
%\end{equation}
%Note that ordering of the sequence of received power levels $S_r, \, r = 1,\dots,h$ is irrelevant in case of capture only receiver.

\subsection{Model Analysis}
\label{subsec:modelanalysis}
The model analysis is carried out by considering the point of view of a tagged node, say node $i$. 
We drop the subscript denoting the tagged node unless required to avoid ambiguity.
If not stated explicitly, it is understood that each variable or quantity refers to the tagged node.

% A backlogged node attempts transmission in a slot with probability $p$.
% When transmitting, a node selects a target \ac{SNIR} $\gamma$.
We follow two schemes to set the transmission parameters:
\begin{itemize}
    \item \emph{Fixed parameter}. Given $n$ nodes in the system, the transmission probability and the \ac{SNIR} threshold are fixed  once and for all as follows:
    \begin{equation}
    \label{eq:p_gamma_starkfixed}
        p^* = 1, \qquad \qquad  \gamma^* = \frac{ 1 }{ a_\gamma n + b_\gamma }
    \end{equation}
%    \begin{equation}
%    \label{eq:lambdastarkfixed}
%        \gamma^* = \frac{ 1 }{ a_\gamma n + b_\gamma }
%    \end{equation}
    \item \emph{Adaptive parameter}. Given that $Q(t) = k$ nodes are backlogged at the beginning of slot $t$, with $1 \le k \le n$, the transmission probability and the \ac{SNIR} threshold are set as follows:
    \begin{equation}
    \label{eq:pstarkadaptive}
        p^*_k = \begin{cases}
         \frac{ 1 }{ k } & \text{ for } 1 \le k < k_c, \\
         1 & \text{ for } k \ge k_c.
    \end{cases}
    \end{equation}
    \begin{equation}
    \label{eq:gammastarkadaptive}
    \gamma^*_k = \begin{cases}
        \gamma_{\text{max}} & \text{ for } 1 \le k < k_c, \\
         \frac{ 1 }{ a_\gamma k + b_\gamma } & \text{ for } k \ge k_c.
    \end{cases}
    \end{equation}
\end{itemize}
It is important to select values of the constants $k_c$, $a_\gamma$ and $b_\gamma$ carefully.
Let us define the sum-rate of the system with $n$ nodes, for given values of $p$ and $\gamma$:
\begin{equation}
\label{eq:sumratexpression}
U(p,\gamma) = \log_2(1+\gamma) \sum_{ h = 0 }^{ k }{ m_h(\gamma) \binom{k}{h} p^k (1-p)^{k-h} } 
\end{equation}
where $m_h(\gamma)$ is the mean number of packets successfully decoded, given that $h$ nodes transmit in the same time slot\footnote{We do not provide analytical expressions for the functions $m_h(\gamma)$. It is actually computationally simpler to get those functions numerically by means of ad-hoc simulations of the \ac{SIC} decoder.}.
% The sum on the right-hand side of \cref{eq:sumratexpression} gives the mean number of packets correctly decoded in the considered time slot.
The sum-rate gives the achieved spectral efficiency of the multiple access channel in bits/s/Hz \cite{Razzaque2022}.

\cref{eq:pstarkadaptive,eq:gammastarkadaptive} provide an asymptotically sharp approximation of the values of $p$ and $\gamma$ that maximize the sum rate, as $k \rightarrow \infty$.\footnote{The development of the formal proofs is lengthy and cannot be included here for space reasons. It can be consulted in \cite{baiocchi2024asymptotic}.}
As a matter of example, for $\epsilon = 0.1$ and $\gamma_{\text{max}} = 31$, it is found that $k_c = 6$, $a_\gamma = 0.39$, and $b_\gamma = 0.78$.

In the rest of this section, we derive the performance metrics in case of \emph{fixed parameters}.
The analysis of the adaptive parameter model is deferred to \cite{internalreport_adaptivetheory_part2} for space reasons.

With a fixed parameter setting, assigned once the overall number of nodes in the system is given, the slot time is fixed to $T^* = T(\gamma^*)$, (see \cref{eq:slottimeduration}).

% \begin{equation}
% T^* = \frac{ L }{ W \log_2(1+\gamma^*) }
% \end{equation}

Let us define the probability distribution of the number of backlogged nodes seen by a tagged node:
\begin{equation}
\label{eq:qPDF}
q_k = \binom{n-1}{k} b^k (1-b)^{n-1-k} \, , \quad k = 0,1,\dots,n-1
\end{equation}
Here $q_k$ is the probability that $k$ nodes, out of the $n-1$ nodes different from the tagged one, are backlogged.
The parameter $b$ is the probability that a node is backlogged at the beginning of a slot.
The probability $b$ is found based on the renewal reward theorem as the ratio of the mean number of slots where the node is backlogged to the sum of the mean number of slots where the nodes is backlogged and the mean number of slots where it is idle:
\begin{equation}
\label{eq:backlogprob}
b = \frac{ 1/p^* }{ 1/p^* + \frac{ 1 }{ 1 - e^{ - \lambda T^* } } } = \frac{ 1 }{ 1 + \frac{ 1 }{ 1 - e^{ - \lambda T^* } } }
\end{equation}
where we have accounted for the choice $p^* = 1$.

% The backlogged probability $b$ cannot attain 1, since it is assumed that nodes have no buffer.
% Hence, a node that completes a transmission attempt must wait at least one slot time, before a new message is handed out from the upper layers to the MAC layer of that node.

\subsubsection{Inter-departure time}
\label{subsubsec:Ytime}
Let $Y$ denote the inter-departure time between two consecutive packets transmitted by the tagged node.
Then we have:
\begin{equation}
\label{eq:Ydefinition}
Y = R + C 
\end{equation}
where $R$ is the idle time of the tagged node and $C$ denotes the contention time, defined as the number of slot times it takes for the tagged node to attempt a transmission, once it becomes backlogged.
The mean of $C$ is given by $\mathrm{E}[ C ] = T^*/p^*$.
% Let $R$ denote the time elapsing since the end of a transmission attempt of the tagged node, until the end of the slot where it becomes backlogged again.
As for $R$, it is distributed geometrically, i.e., it is 
% The Laplace transform of the \ac{PDF} of $R$ is
\begin{equation}
\mathcal{P}( R = j T^* ) = \left( e^{ - \lambda T^* } \right)^{j-1} \left( 1 - e^{ - \lambda T^* } \right) \, , \quad j \ge 1.
% \varphi_R(s) = \frac{  - 1 }{ e^{ \lambda (s+\lambda) } - 1 }
\end{equation}
and the mean is $\mathrm{E}[ R ] = T^*/( 1-e^{ - \lambda T^* } )$.

\subsubsection{Success Probability and \ac{CBR}}
% \label{subsubsec:sucessprob_throughput}

Let $\tau = b p^*$ be the probability that the tagged node transmits in a given slot.
The probability of success is found as follows:
\begin{equation}
%\label{ }
P_s = \frac{ 1 }{ n \tau } \sum_{ h = 1 }^{ n }{ m_h(\gamma) \binom{n}{h} \tau^h (1 - \tau)^{n-h} }
\end{equation}

The \ac{CBR} is the mean fraction of time that the tagged node senses the channel as busy.
It is found simply as:
\begin{equation}
%\label{ }
\text{CBR} = 1 - ( 1 - b p^* )^n
\end{equation}

\subsubsection{Throughput}

Given the generation rate $\lambda$ of messages at a node, there are two sources of message loss: (i) dropping of arriving messages when the tagged node is busy in contention or transmission; (ii) failed decoding.
The mean rate of messages sent on air by a node is $1/\mathrm{E}[Y]$, i.e., the mean inter-departure rate.
The net throughput in messages per unit time is therefore given by $\Theta = P_s/\mathrm{E}[Y]$.
%\begin{equation}
%    \Theta = \frac{ P_s }{ \mathrm{E}[Y] } 
%\end{equation}
The throughput in $\si{\bit\per\second}$ can be obtained by considering the message payload $L$, i.e., it is $\Theta_{\text{bps}} =L \Theta$.

\subsubsection{Access Delay}
% \label{subsubsec:AccessDelay}

The access delay $D$ is defined as the interval between the arrival time of a message and the completion of the transmission time of that message.
The access delay is $D = V+C$, where $V$ is the time elapsing since the \emph{last} arrival within a time slot and the end of that time slot.
The mean access delay is given by:
\begin{equation}
\mathrm{E}[ D ] = \frac{ 1 }{ \lambda } - \frac{ T^* }{ e^{ \lambda T^* } - 1 } + \frac{ T^* }{ p^* }
\end{equation}
It is easily recognized that $T^*/p^* < \mathrm{E}[ D ] \le T^*/p^* +T^*/2$ as $\lambda$ grows from 0 to $\infty$.

\subsubsection{Age of Information}

The \ac{AoI} is defined as the age of data stored at the \ac{BS} for the tagged node.
It is akin to the excess random variable associated to the random variable $Z$, defined as the time elapsing between two successive successful reception of messages coming from the tagged node.
Recalling that $Y$ denotes the inter-departure time of packets from the tagged node, we have
\begin{equation}
    Z = \sum_{ i = 1 }^{ N }{ Y^{(i)} }
\end{equation}
where $N$ is the number of attempts required to achieve a successful message delivery.
The random variable $N$ is Geometrically distributed, with $\mathcal{P}( N = k ) = (1-P_s)^{k-1} P_s, \; k \ge 1$.

Accounting also for the mean access delay, the mean \ac{AoI} can be written as
\begin{equation}
    \mathrm{E}[ A ] = \mathrm{E}[ D ] + \frac{ \mathrm{E}[ Z^2 ] }{ 2 \, \mathrm{E}[ Z ] } =  \mathrm{E}[D] + \frac{ \mathrm{E}[ Y^2 ] }{ 2 \mathrm{E}[Y] } + \mathrm{E}[ Y ] \left( \frac{ 1 }{ P_s } - 1 \right)
\end{equation}
We recall that $Y = R+C$.
Since $p^* = 1$, it is $C = T^*$.
As for $R$, it has a Geometric \ac{PDF}.
Hence
\begin{align}
	&\mathrm{E}[ Y ] = T^* + \frac{ T^* }{ 1 - e^{ - \lambda T^* } }  \\
	&\mathrm{E}[ Y^2 ] = \left( T^* \right)^2 \left[ 1 + \frac{ 3 - e^{ - \lambda T^* }}{ \left( 1 - e^{ - \lambda T^* } \right)^2 } \right]
\end{align}

% -------------- Section end marker --------------
%                _       _
%               ( )_    ( )
%    ___  _   _ | ,_)   | |__     __   _ __   __
%  /'___)( ) ( )| |     |  _ `\ /'__`\( '__)/'__`\
% ( (___ | (_) || |_    | | | |(  ___/| |  (  ___/
% `\____)`\___/'`\__)   (_) (_)`\____)(_)  `\____)
%
% -------------- Section end marker --------------

\section{Numerical Evaluations}
\label{sec:num_eval}

The numerical evaluations are done in the following way for each scheme.
\begin{itemize}
    \item \emph{Fixed parameters}:
    In case of the fixed parameter model, we only used the analytical expressions given in \cref{sec:model_and_analysis} to derive the performance metrics as there exists no assumption and the system is simple enough.
    The slot time, along with transmission probability, and the target \ac{SNIR}, is fixed to a constant number given by \cref{eq:p_gamma_starkfixed}.
    \item \emph{Adaptive parameters}:
    In case of the adaptive parameter model, only simulations are done.
    For simulations, we fully consider the evolution of the system.
    Each node evolves according to a two-state Markov chain.
    A node is idle until a new message is generated.
    At the end of the slot where a new message is generated, the node transitions to the active state, where it contends for the channel.
    While active, it transmits with probability $p_k$ in a slot, if $k$ nodes are backlogged at the beginning of that slot.
    Immediately after having transmitted, the node moves back to the idle state.
    The slot time in the simulation is set to $T_k$ if $k$ nodes are backlogged at the beginning of the slot.
    The time-varying size of slot times gives rise to a complex interplay between nodes, since the more nodes are active in one slot, the longer its duration, the higher the probability that a new message arrives at those nodes that are not active in that slot.
\end{itemize}

In the results presented, we consistently set the number of nodes to $n = 50$, and the packet size to $L = \SI{500}{\byte}$, while varying the mean message generation time $S = 1/\lambda$ between 1 ms and 1000 ms.
The \emph{fixed parameter} results are displayed with a solid blue line, while the \emph{adaptive parameter} results are displayed with a dashed circle blue line.
The other numerical values of the main system parameters include $P_N = -107$ dBm and $W = 1$ MHz.

\subsection{Light and heavy traffic regimes}
\label{subsec:lightheavyregimes}

Numerical results are plotted as a function of the mean message generation time, $S = 1/\lambda$.
We consider a quite stretched range of $S$ values, to highlight the existence of two different operational regimes of the system.
Low values of $S$, lying on the left side of the $x$ axis of each plot is referred as a heavy traffic regime, corresponding to nodes generating new messages very frequently, i.e., with mean generation time smaller than the average time slot size.
On the opposite side of $x$-axis, large values of $S$ correspond to the light traffic regime, where nodes generate new messages infrequently, imposing a light load on the channel, given that message generation times are much bigger than the average slot time.
In between these two regions, there exists a region that can be called a transition region.

\subsection{\ac{PDR} and access delay}
\label{subsec:PDRandaccessdelay}

\begin{figure}[t]
\centering
\includegraphics[width=0.4\textwidth]{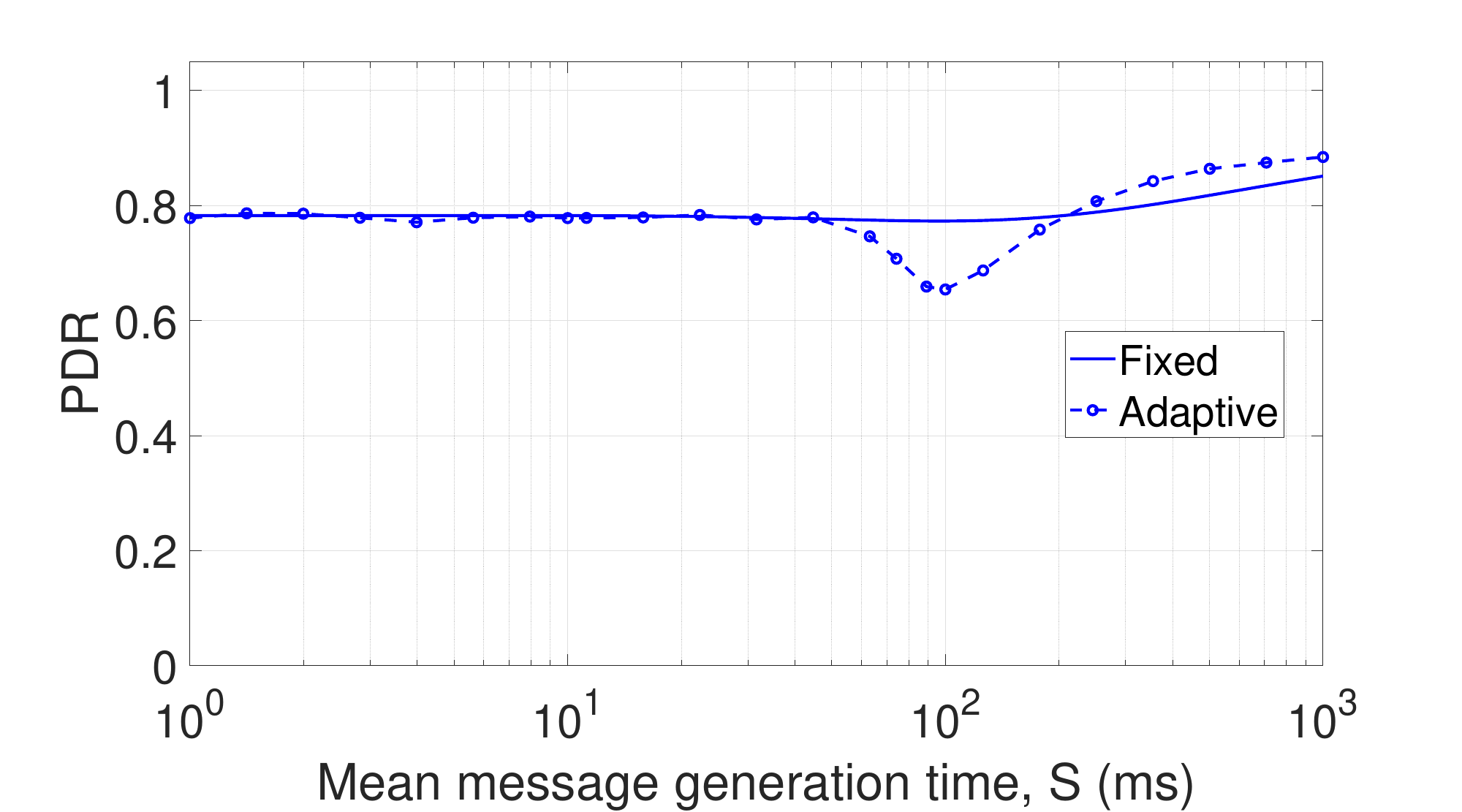}
\caption{Packet delivery ratio as a function of $S$.}
\label{fig:PDR_S}
\end{figure}

\ac{PDR} is plotted in \cref{fig:PDR_S} as a function of $S$.
The \ac{PDR} measures the mean fraction of packets that are delivered successfully to the \ac{BS}.
An high level of \ac{PDR} is achieved in the heavy traffic regime, which gives evidence of the effectiveness of \ac{SIC} in dealing with a large number of backlogged nodes.
The \ac{PDR} in heavy traffic regime is the same for both \emph{fixed} and \emph{adaptive} schemes.
This is based on the fact that the optimal parameters in the case of \emph{fixed parameter} scheme are set to a maximum number of nodes present in the system but the number of backlogged nodes never crosses this limit, hence every node that is backlogged is allowed to transmit successfully, but with low data rates as compared to \emph{adaptive} scheme as the $\gamma$ is less for \emph{fixed parameter} scheme.

In the low traffic regime, the number of backlogged nodes is typically below $k_c$ and hence following the \emph{adaptive parameter} scheme the transmission probability is set to $1/k$, which entails that the expected number of transmitting nodes is 1.
If more nodes transmit simultaneously, failure is most probable, given that $\gamma_k = \gamma_{\text{max}}$ in this region.
The system dynamics resemble those of classic \ac{SA}, wherein the success probability collapses as the load increases (i.e., $S$ decreases in our scenario).
In contrast to classic \ac{SA}, in our scenario, as we approach the transition region, the number of backlogged nodes grows, and \ac{SIC} is increasingly triggered when $k > k_c$.
Then, we transition into the heavy traffic regime.

The down notch seen in the \ac{PDR} plot is reminiscent of the performance drop of classic \ac{SA}.
When $S$ decreases from the right of the plot, the load on the system grows, and the probability of failing decoding with $\gamma = \gamma_{\text{max}}$ grows.
While classic \ac{SA} throughput collapses as the load further increases, here the adaptation of the parameters $p$ and $\gamma$ restores high \ac{PDR} values, to the cost of slowing down transmission rate (longer time slots are used).
In case of \emph{fixed parameter} scheme, we don't see any down notch as the probability of failing decoding doesn't drop due to lower $\gamma^*$.

\begin{figure}[t]
\centering
\includegraphics[width=0.4\textwidth]{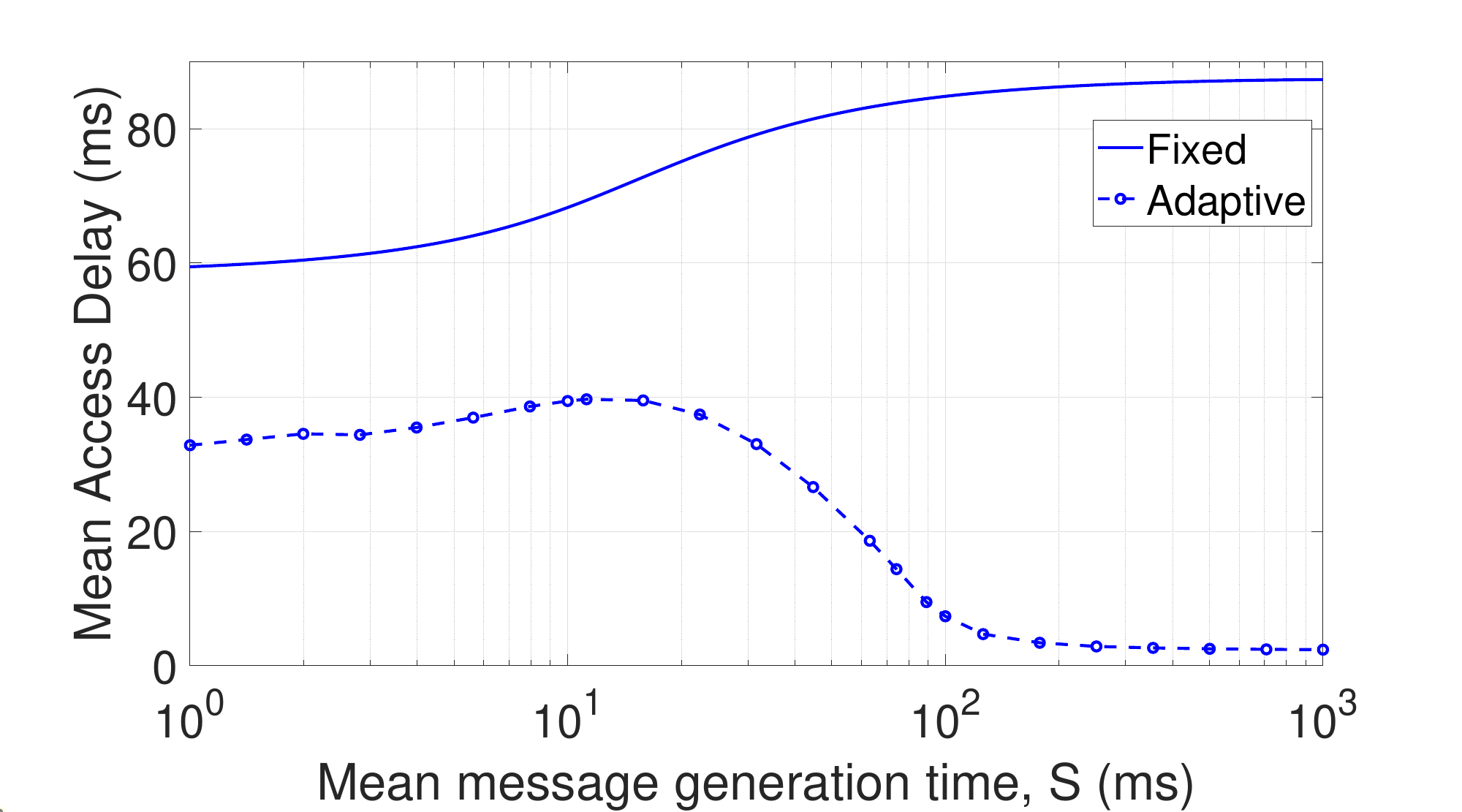}
\caption{Mean access delay as a function of $S$.}
\label{fig:EAD_S}
\end{figure}

The mean access delay is shown in \cref{fig:EAD_S} as a function of mean message generation time $S$.
The mean access delay is defined as the duration starting from when the message intended for transmission arrives at the node until the transmission of that message is completed.

In the case of a \emph{fixed parameter} scheme, the mean access delay is very high as compared to an \emph{adaptive parameter} scheme in a heavy traffic regime, which is based on the fact that the parameter $\gamma$ is set based on overall nodes present in the systems i.e. $n$ instead of a backlogged number of nodes $k$. 
Hence, the transmission time $T$ differs accordingly for both schemes.

Smaller access delays are seen in the light traffic regime for the \emph{adaptive parameter} scheme, mainly because of the much smaller transmission time, even if using $p_k <1$ introduces a non-null contention time, i.e., on average $1/p_k$ time slots are required before transmission is attempted.
While it gets even worse in the light traffic regime for \emph{fixed parameter} scheme.
The intermediate peak in case of an \emph{adaptive parameter} scheme, stems from the adverse effect of the transition region, where the system oscillates between non-null contention time and immediate transmission ($p_k = 1$), but with a large slot time.

\subsection{Throughput and \ac{AoI}}
\label{subsec:thruenergyAoI}

Throughput and mean \ac{AoI} are presented in this section. 
Throughput measures the mean delivered bit rate or, if normalized, the mean fraction of generated messages that are successfully delivered to the base station.
\ac{AoI} is the well-known metric \cite{Yates2021}, referred to the age of the last current update data generated by each node and stored in the collecting base station.

\subsubsection{Throughput}

\begin{figure}[t]
\centering
\includegraphics[width=0.4\textwidth]{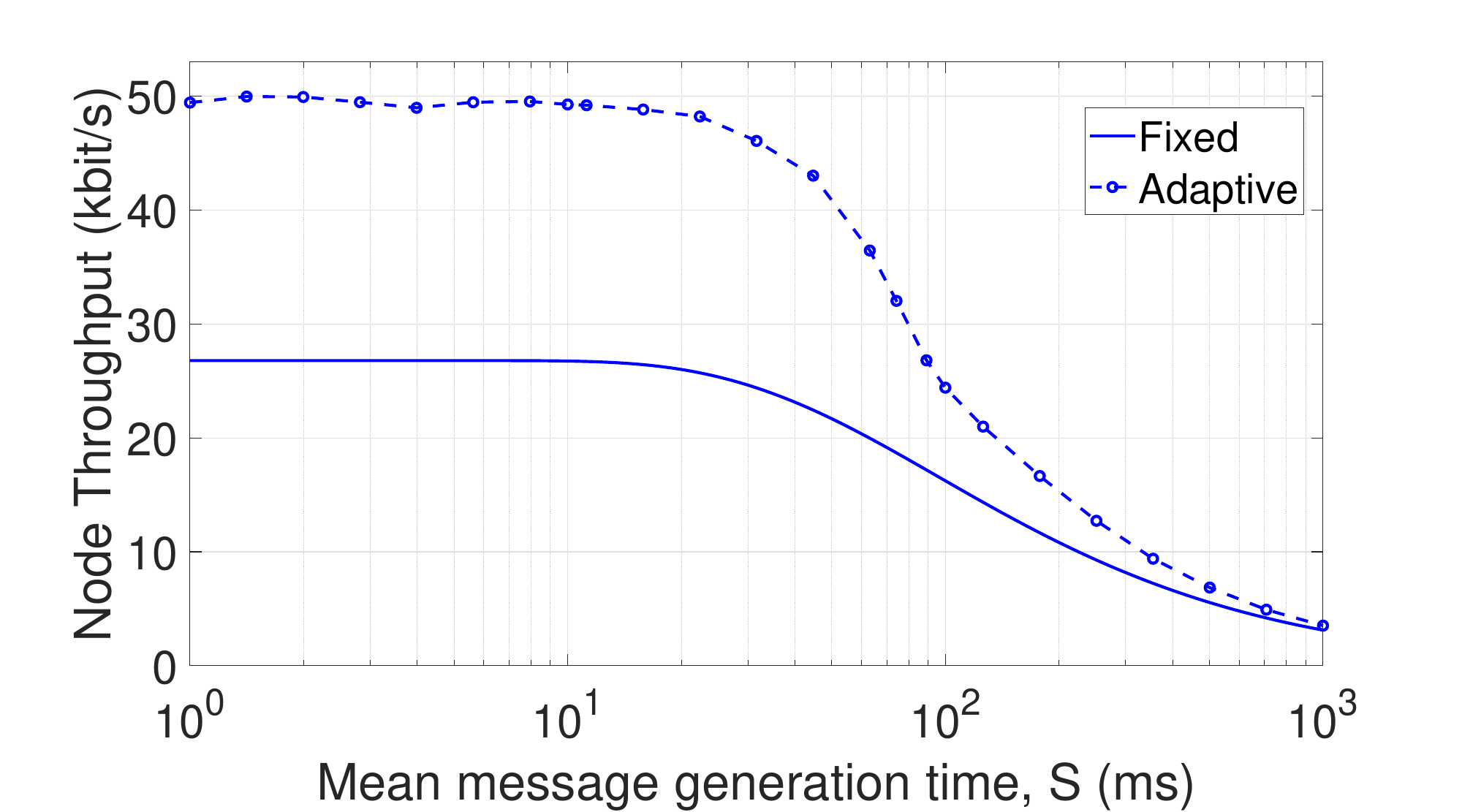}
\caption{Throughput in kilo-bit per second as a function of $S$.}
\label{fig:thru_vs_S}
\end{figure}

\begin{figure}[t]
\centering
\includegraphics[width=0.4\textwidth]{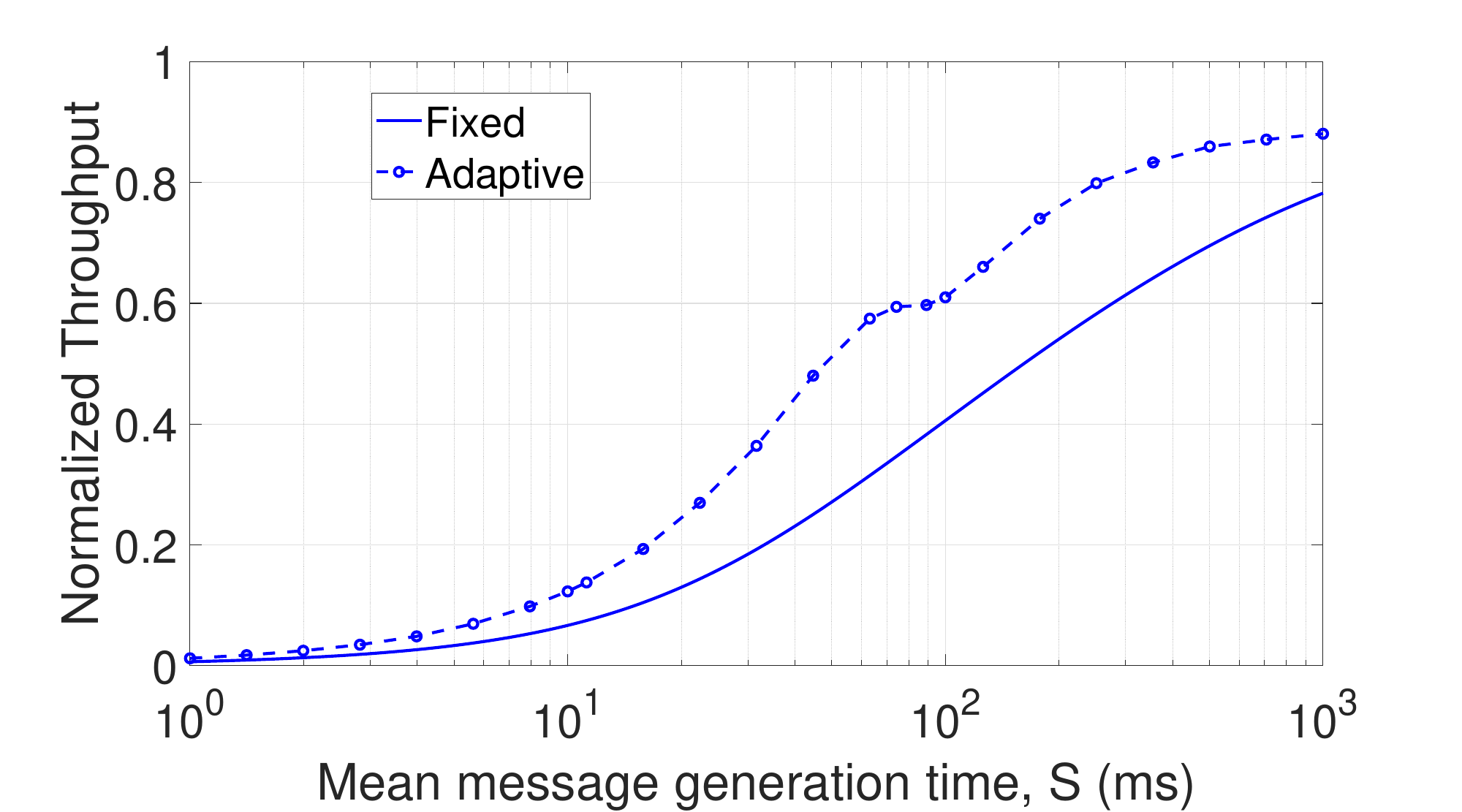}
\caption{Normalized throughput as a function of $S$.}
\label{fig:nthru_vs_S}
\end{figure}

The node throughput and normalized throughput are shown in \cref{fig:thru_vs_S,fig:nthru_vs_S} respectively, as a function of mean message generation time $S$.
It is recalled that node throughput in \cref{fig:thru_vs_S} is the mean carried bit rate of messages delivered to the base station, while in \cref{fig:nthru_vs_S} it is normalized with respect to message generation rate $\lambda = 1/S$.

The node throughput in \cref{fig:thru_vs_S} saturates when the system is pushed into the heavy-traffic regime.
The \emph{adaptive parameter} multiple access scheme appears to scale robustly, with no collapse as the rate of generation of update messages increases in the limit for $S \rightarrow 0$.
Although \emph{fixed parameter} multiple access scales with lower throughput compared to the \emph{adaptive parameter}, due to low $\gamma^*$, which results in much longer transmission slot times, and consequently throughput decreases accordingly.
As $S$ increases, after the transition region, the node throughput falls for both schemes, as expected, given the diminishing generation rate of new update messages.

\cref{fig:nthru_vs_S} shows that normalized throughput increases as the message generation rate increases, falling to negligible values for very low values of $S$.
This behavior can be understood by analyzing the sources of message loss.
There are two sources of loss.
First, messages offered by upper layers to node MAC entity are discarded, if the MAC entity is engaged in contention or in transmission.
Second, messages that are not decoded successfully, because of failure of \ac{SIC}, are lost as well.
The normalized throughput performance in the heavy traffic regime is primarily influenced by the first source of message loss.
Both sources of loss have comparable impacts in the transition region between heavy and light traffic regimes.
Packet loss is dominated by residual decoding errors when moving to the light traffic regime.
In case of a \emph{fixed parameter} scheme, the normalized throughput is much lower than the \emph{adaptive parameter} scheme as we move from a heavy traffic regime to a low traffic regime, which refers to much higher failing decoding as the selected \ac{SNIR} threshold is much lower in \emph{fixed parameter} scheme.

Summing up, the two plots of throughput suggest that the highest possible throughput rate (saturation) is achieved in an \emph{adaptive parameter} scheme under heavy traffic i.e. typical \ac{IoT} operating regime, which however entails that most of the generated messages are discarded before any transmission attempt occurs.
The highest efficiency, indicated by a large \emph{normalized} throughput, is achieved under the light traffic regime in case of an \emph{adaptive parameter} scheme.
However, the throughput rate is relatively low in this regime.

\subsubsection{AoI}

\begin{figure}[t]
\centering
\includegraphics[width=0.4\textwidth]{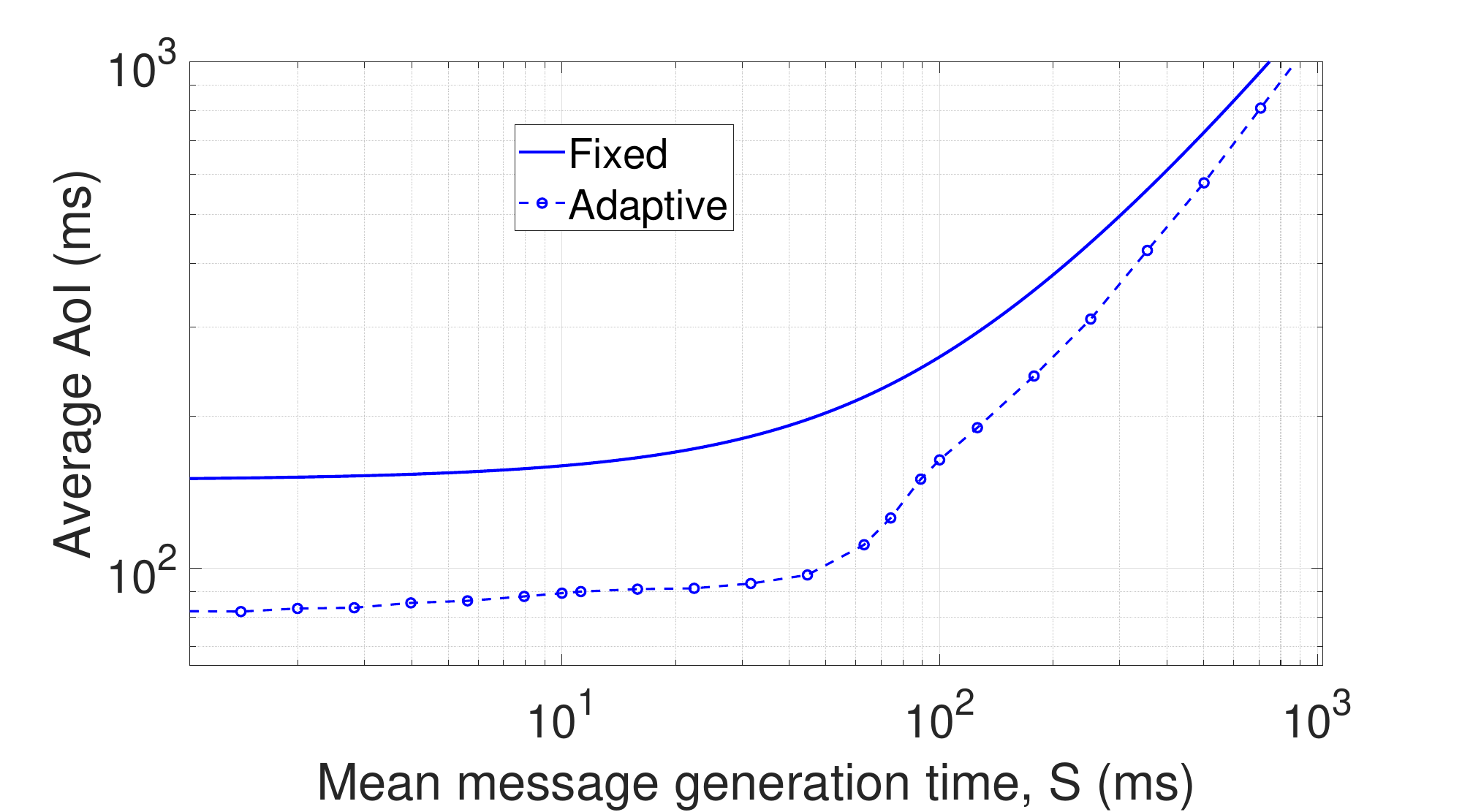}
\caption{Mean AoI as a function of $S$}
\label{fig:aoi_vs_S}
\end{figure}

The mean \ac{AoI} is shown in \cref{fig:aoi_vs_S} as a function of mean message generation time $S$.
In the light-traffic regime, the mean age of information is very high for both schemes, as we generate fewer and fewer updates.
As we move to a heavy-traffic regime, the \ac{AoI} gets low, since update messages are generated more frequently and \ac{SIC} helps relieving the congestion on the channel.
In case of an \emph{adaptive parameter} scheme, the age of information is lower than the \emph{fixed parameter} scheme, due to less mean access delay.
As we move from a low-traffic regime to a heavy traffic regime this difference almost doubles.
The optimal region in terms of the mean age of the information is the heavy-traffic regime with an \emph{adaptive parameter} scheme.

\section{Conclusions}
\label{sec:conclusions}

We have presented a model for a random multiple access system where nodes send messages to a central \ac{BS}, whose receiver exploits \ac{SIC}.
The model is targeted to \ac{IoT} applications.
The results shown answer the question of whether setting up an \emph{adaptive} multiple access scheme buys enough performance gain to warrant its added complexity.
It is apparent that significant performance gains are obtained, especially in terms of delay and mean \ac{AoI}, which are key metrics in \ac{IoT} scenarios.
One can gain in terms of high throughput, low access delay, and lower \ac{AoI} using the \emph{adaptive parameter} scheme over the \emph{fixed parameter} approach.
These prominent gains call for the designing of complex adaptive systems instead of fixed systems for \ac{IoT} networks.
Further work should address the identification of practical algorithms to estimate the number of backlogged nodes and practical coding schemes for \ac{SIC}, to highlight the price to be paid for a practical implementation of the general \ac{SIC} receiver considered in this paper.

\section*{Acknowledgement}

This work was partially supported by the European Union under the Italian National Recovery and Resilience Plan (NRRP) of Next Generation EU, partnership on ``Telecommunications of the Future'' (PE00000001 - program ``RESTART'').

\end{document}